\documentclass[pdflatex,sn-chicago]{sn-jnl}

\usepackage{graphicx}%
\usepackage{multirow}%
\usepackage{amsmath,amssymb,amsfonts}%
\usepackage{amsthm}%
\usepackage{mathrsfs}%
\usepackage[title]{appendix}%
\usepackage{xcolor}%
\usepackage{textcomp}%
\usepackage{manyfoot}%
\usepackage{booktabs}%
\usepackage{algorithm}%
\usepackage{algorithmicx}%
\usepackage{algpseudocode}%
\usepackage{listings}%
\usepackage{natbib}

\theoremstyle{thmstyleone}%
\theoremstyle{thmstyletwo}%

\theoremstyle{thmstylethree}%

\begin{document}

\title[5 Days, 5 Stories]{5 Days, 5 Stories: Using Technology to Promote Empathy in the Workplace}


\author*[1]{\fnm{Russell} \sur{Beale}}\email{r.beale@bham.ac.uk}

\author[1]{\fnm{Eugenia} \sur{Sergueeva}}\email{eugenia@azureindigo.com}

\affil*[1]{\orgdiv{School of Computer Science}, \orgname{University of Birmimgham}, \orgaddress{\street{Edgbaston}, \city{Birmingham}, \postcode{B15 2TT}, \country{UK}}}


\abstract{Empathy is widely recognized as a vital attribute for effective collaboration and communication in the workplace, yet developing empathic skills and fostering it among colleagues remains a challenge. This study explores the potential of a collaborative digital storytelling platform — In Your Shoes — designed to promote empathic listening and interpersonal understanding through the structured exchange of personal narratives. A one-week intervention was conducted with employees from multiple organizations using the platform. Employing a mixed methods approach, we assessed quantitative changes in empathy using the Empathy Quotient (EQ) and qualitatively analyzed participant experiences through grounded theory. While quantitative analysis revealed no statistically significant shift in dispositional empathy, qualitative findings suggested the tool facilitated situational empathy, prompted self-reflection, improved emotional resonance, and enhanced workplace relationships. Participants reported feelings of psychological safety, connection, and, in some cases, therapeutic benefits from sharing and responding to stories. These results highlight the promise of asynchronous, structured narrative-based digital tools for supporting empathic engagement in professional settings, offering insights for the design of emotionally intelligent workplace technologies.}




\maketitle

\section{Introduction}\label{sec1}
Empathy --- vicariously experiencing the feelings of others --- is an important characteristic of human experience, communication and understanding. Workplaces are communities of people, and enhancing understanding and encouraging empathy within them has therefore garnered significant attention in organizational behavior and leadership studies.  It is recognized as a pivotal component in  effective leadership, enhancing team dynamics \citep{HBR2023}, and improving overall organizational performance.  Empathetic leadership involves understanding employees' perspectives and emotions, which contributes to building trust and improving communication, creating supportive environments that enhance employee satisfaction and productivity \citep{FranklinCovey2024, gentry2007, Gentry2014}

Organizations have implemented training programs aimed at developing empathy among employees and leaders \citep{RocheMartin2023}. These programs often include role-playing exercises, active listening workshops, and perspective-taking activities designed to cultivate emotional intelligence and empathetic communication skills. Studies suggest a positive correlation between empathetic leadership and employee performance. Employees are more likely to be motivated and demonstrate commitment to organizational goals when they perceive their leaders as empathetic. 

Previous studies have also shown that a physician’s or a therapist’s ability to express their empathy positively impacts both themselves and their patients \citep{Hojat2007}. Empathy is also known to facilitate prosocial behaviour \citep{Eisenberg1987} and increase one’s attention to the wellbeing of others \citep{Mehrabian1972}.

Previous research has established that improving listening and interpersonal skills, discussing shared experiences, perspective taking, and self-disclosure aid in feeling empathic towards others. Consequently, this encourages individuals to be more concerned about other people’s hardships, be more perceptive to the feelings of others, and want to help, share, and co-operate. Based on these principles, a high-fidelity prototype of a website for exchanging personal stories \textit{In Your Shoes} has been developed. This platform aims at providing a safe, amicable environment, where Storytellers --- those who share their experience ---- would be heard, and Listeners  --- those who read and engage with a story--- would be able to learn something new about their colleagues and understand them better. Two research questions shaped the framework of this study:
\begin{enumerate}
    \item What effect has this online tool for exchanging personal stories on the participants’ empathy in the workplace environment?
    \item What is the participants’ experience of using this tool?
\end{enumerate}

\section{Defining Empathy}

\subsection{Cognitive and Affective Empathy}
According to the Merriam-Webster dictionary, empathy is ``the action of understanding, being aware of, being sensitive to, and vicariously experiencing the feelings, thoughts, and experience of another of either the past or present without having the feelings, thoughts, and experience fully communicated in an objectively explicit manner'' \citep{MerriamWebster}.   Another definition, proposed by Encyclopædia Britannica, suggests that empathy is ``the ability to imagine oneself in another’s place and understand the other’s feelings, desires, ideas, and actions'' \citep{Britannica}.

Both of these definitions combine two aspects: an ability to intellectually understand other people’s emotions and the origins of them (their cognitive nature) and an ability to associate with other people’s feelings, for example, by recalling a similar experience (their affective nature).

On one hand, a large body of literature conceptualises empathy as a cognitive attribute (for example, \citet{hojat2013empathy}, \citet{hogan1969},  \citet{basch1983}, \citet{truax2007}). Hojat defines empathy as the ability to understand the “experiences, concerns and perspectives of another person, combined with a capacity to communicate this understanding” \citep{hojat2009approaches}. Cognitive empathy is also often called perspective taking or role taking. 

Another group of researchers claim that empathy is predominantly an affective (emotional) attribute, that involves feeling and appreciating the emotional state of others in a visceral way (e.g. \citet{Mehrabian1972}, \citet{Eisenberg1987}). As opposed to cognitive empathy, affective empathy, or affective role taking, involves matching the emotions of others. In their analysis, Eisenberg and Miller identify affective empathy as a “state that stems from the apprehension of another’s emotional state or condition, and that is congruent with it”. \citet{shapiro2002} suggests that feeling empathic implies sensing other people’s feelings: “empathy is more than intellectual understanding or cognitive analysis”. Furthermore, as noted by Eisenberg and Miller, affective empathy is also conceptualised differently by different theorists. First, emotional empathy can be defined as “vicarious matching of another’s affective state” or feeling the same emotion as another person \citep{Feshbach1968, vankleef2008}. Second, some researchers suggest that affective empathy is a concern about another person’s state \citep{batson1981}: they claim that empathic people sympathise with others. Finally, yet others argue for a combination of both affective matching and sympathy when defining empathy \citep{Mehrabian1972}. 

Many theorists argue that empathy is an amalgamation of cognitive and affective concepts (e.g. \citet{BaronCohen2004,Davis1980}). They claim that these two aspects of empathy form an interdependent system and cannot be studied separately \citep{Feshbach1968}.

\subsection{Empathy as a Trait, a State, or a Process}
Several studies refer to empathy as a stable human characteristic or general ability (e.g. \citet{Davis1980,Mehrabian1972,hogan1969,dymond1950}). Hogan suggests the term \textit{empathic dispositions} to describe this concept. Similarly, Davis describes it as dispositional empathy. This approach assumes that some people are more empathic than others either innately or by developing this quality in them. Duan and Hill recommend this conceptualisation for those researchers who wish to examine inter-individual variations of empathy levels \citep{Duan1996}. Additionally, this assumption can support theories connecting empathy to other personal traits \citep{Feshbach1968} or moral practices such as altruism \citep{batson1981}.

Other studies define empathy as a situation-specific reaction, or a state \citep{Feshbach1968}. Theorists who adhere to this theory argue that this state, or vicarious empathic experience, is not determined solely by intelligence or inherent empathy levels, but varies with the emotional reaction to a specific situation \citep{Feshbach1968}. Duan and Hill point out that this approach allows for studying the effect of training on one’s empathy levels. An alternative approach adopted by theorists is to view empathy as a multi-phased process (for example, \citet{rogers_empathic_1975,barrettlennard1981}). These researchers analyse the stages an individual goes through when they experience empathy: from relating to one’s feelings, communicating their empathic state in response, and having their response acknowledged by that person. There are various models describing this process, and they all chiefly aim at studying empathy in the counselling or psychotherapeutic contexts. 

\subsection{Empathy and Sympathy}
Theorists often consider empathy in connection with related constructs like sympathy. Despite a large volume of studies on empathy, there exists some confusion in using these terms. As Spiro notes: "It is difficult to distinguish empathy from sympathy: where empathy feels `I am you,' sympathy may mean `I want to help you'" \citep{Spiro1992}. Although sometimes used interchangeably, the literature generally treats them separately (e.g. \citet{Hojat2007,Mercer2002}). Hojat states that sympathy is "predominantly an affective or emotional attribute that involves intense feelings of a patient’s pain and suffering" \citep{Hojat2007}. Eisenberg and Miller also regard sympathy as an affective response to another’s internal state \citep{Eisenberg1987}, arguing that it is driven by concern for the other person.

Unlike Hojat or Eisenberg and Miller, others like \citet{mead1934} argue that sympathy is a cognitive characteristic that allows one to understand other people’s emotions. Moreover, Shapiro claims that there is a risk of having too much empathy, that empathy “can be an overwhelming and emotionally burdensome experience” \citep{shapiro2002}. It is worth noting, however, that this conclusion derives from the definition of empathy as an affective characteristic.  \citet{hojat2009approaches} notes empathy and sympathy should only be distinguished in a clinical setting in the context of a doctor-patient relationship, and argues that it is not essential to separate these concepts when referring to a social context: both empathy and sympathy may lead to prosocial behaviour. He highlights that the only difference is the triggering factor for that behaviour: self-oriented for sympathy versus other- oriented for empathy. Eisenberg and Miller also argue that both empathy and sympathy may lead to prosocial and altruistic behaviour. However, they note that both of these constructs are not self-oriented. 

\section{The Importance of Empathy }
Numerous studies have highlighted the importance of empathy in various settings \citep{Duan1996, Gentry2014}. As well as more conventional commercial settings, empathy is central to the doctor-patient relationship \citep{Hojat2007,Hojat2009,suchman1997}, client-centred therapeutic approaches \citep{rogers_empathic_1975}, treatment of sex offenders \citep{mcgrath2010} and domestic abusers \citep{mead1934}, violence prevention programs \citep{grossman1997}, and anger management courses \citep{wilkes2002}. Research has shown that a physician’s ability to demonstrate empathy positively affects patient satisfaction \citep{kim2004} and may be linked to the professional satisfaction of doctors \citep{suchman1993}. Shapiro notes that empathy makes "the practice of medicine more rewarding, more interesting, less frustrating, and more pleasurable" \citep{shapiro2002}. Gentry et al. assert that high empathy levels might correlate with better job performance \citep{gentry2007}. Some theorists claim that in clinical settings, excessive sympathy may be detrimental and lead to emotional exhaustion, whereas the cognitive nature of empathy contributes to workplace satisfaction \citep{Hojat2009}.

 High levels of empathy are critical in clinical settings, yet research indicates that empathy often declines over time. Spiro believes that such factors as “isolation, long hours of service, chronic lack of sleep” \citep{Spiro1992} are detrimental to empathic capacity.  In studies of medical students, Hojat found that factors like "fear of making mistakes, a demanding curriculum, time pressure, sleep loss, and a hostile environment" contribute to declining empathy \citep{Hojat2009}. This decline underscores the need for interventions aimed at sustaining or enhancing empathy. 

\subsection{Empathy Training}
\subsubsection{Interpersonal Training}
Numerous studies have reported that empathy can be trained \citep{DasGupta2004,Hojat2007,shapiro2002}. Hojat suggests methods such as improving interpersonal skills, audio or video-taping patient encounters, exposure to role models, role playing, shadowing a patient, hospitalization experiences, studying literature and the arts, narrative skills development, theatrical performances, and the Balint method \citep{Hojat2009}. While some methods are specific to clinical settings, others have been adapted for different fields. For example, role playing can help Human-Computer Interaction specialists develop empathy for their users by, for example,  simulating conditions such as sensory loss by wearing earplugs. 

Effective listening is crucial for understanding others \citep{gentry2007}. Psychologists suggest that actively listening for the underlying message and paying attention to nonverbal cues helps one to become an empathic listener. Active listening, which involves focusing on the speaker, withholding judgment, and clarifying uncertainties, fosters empathic engagement. Reflective listening, which entails paraphrasing the speaker’s message, further reinforces understanding and trust. These strategies help create a safe environment where individuals feel heard, thereby promoting empathy. Reflective speaking is in the heart of the personcentred psychotherapy and allows the speaker to explore their feelings further and demonstrate them that they are being heard. Both active and reflective listening are strategies of communicating empathy. By paying attention, reflecting, and clarifying, listeners create a safe environment and encourage opening up. Moreover, it has been shown that empathic listening can mitigate social conflicts and raise positive emotions \citep{seehausen2012}.

Interpersonal, or social, skills are vital for effective communication and relationship building. Improving these skills is considered a prerequisite for conveying empathy \citep{hogan1969}. In his major study, Hogan reports that individuals who score high in his empathy scale are more “socially acute and sensitive to nuances in interpersonal relationships”  than those, who score low. 
Models for training social skills, such as those proposed by \citet{suchman1997}, emphasize recognising, understanding, and reflecting others’ feelings.  Having a common experience with another person allows for greater resonance with them and increases chances of feeling empathic towards that person . Spiro suggests that “conversations about experiences, discussions of patients and their human stories” help practitioners to cultivate empathy towards their patients. Therefore, other people’s personal stories may be a powerful way of understanding them better through unveiling shared interests, beliefs, and values \citep{Spiro1992}. Exchanging personal narratives enables individuals to discover shared values and interests, facilitating deeper understanding and connection.

\subsubsection{Computer-based Support}
Whilst training programs and awareness-raising are typical approaches to encouraging an empathetic workplace, there are a number of computer-based tools specifically designed or adapted to enhance empathy, often leveraging digital storytelling, emotional analytics, immersive experiences, or communication facilitation.  Embodied Labs \citep{EmbodiedLabs2024} provides immersive virtual environments that simulate age-related challenges, whilst AR apps that overlay digital information or scenarios onto the real world, allowing users to explore empathy-building situations interactively such as ARTE’s "Notes on Blindness" \citep{NotesOnBlindness2024} which  provides an experiential understanding of living without sight, thus enhancing empathy.  The Chrome web browser has an extension that allows developers to see how their sites look for people with typical visual deficiencies \citep{osmani_emulate_2020}.  There are a range of empathy training applications designed explicitly to teach empathy skills through targeted exercises and reflections: \citet{Empatico2024} connects classrooms globally, allowing children and educators to foster empathy through cross-cultural interactions.

Chatbots and conversational agents can be used to practice empathetic conversations in a safe, simulated environment: \citet{Woebot2024} is an AI chatbot designed with cognitive-behavioral principles to practice emotional support and empathy-driven conversations.  There are serious games explicitly designed to foster empathy by putting players in emotionally engaging situations or roles, such as  "That Dragon, Cancer" \citep{ThatDragonCancer2016}, designed to evoke empathy towards families experiencing terminal illness.

Of more direct relevance to our research, digital storytelling platforms encourage personal storytelling, facilitating deeper empathy through shared emotional experiences.  For example, \citet{StoryCorps2024} provides an online platform for people to share deeply personal stories to build connections and empathy. Online reflective writing tools (who remembers blogs?) can encourage reflective journaling or narrative sharing, enhancing users' emotional self-awareness and empathy towards others e.g. \citet{Penzu2024}.

\section{Measuring Empathy}
Various methods exist for measuring empathy, with distinctions made between situational empathy and dispositional empathy. Situational empathy is assessed in real-time using observational or physiological methods, while dispositional empathy is measured through self-report questionnaires. Common instruments include the Questionnaire Measure of Emotional Empathy (QMEE) \citep{Mehrabian1972}, the Interpersonal Reactivity Index (IRI) \citep{Davis1980}, Hogan’s Empathy Scale \citep{hogan1969}, and the Empathy Quotient (EQ) \citep{BaronCohen2004}. Each instrument has its limitations, and the choice of tool depends on the specific aspects of empathy under investigation.  These tools reflect the heterogeneity of the definitions; they have their limitations and measure different aspects of empathy. For example, the QMEE is designed to measure only the emotional aspect, whilst the Hogan’s Empathy Scale is based on the concept of empathy as an exclusively cognitive characteristic. Davis's Interpersonal Reactivity Index questionnaire taps both affective and cognitive empathy. Even though all these instruments are widely utilised, theorists disagree on which of them provides the best measurement. 

Hogan’s Empathy Scale (1969) consists of 64 items and analyses four factors: “social self-confidence, even temperedness, sensitivity, and nonconformity” \citep{hogan1969}. But \citet{BaronCohen2004} argue that only sensitivity is directly related to empathy, whilst \citet{Davis1980} suggests that this scale is more suitable for measuring social skills instead. This questionnaire does contains questions many would consider probably irrelevant to empathy, such as “I prefer a shower to a tub bath” or “I think I would like to belong to a singing club”, which make theorists question the validity of the scale. 

The Questionnaire Measure of Emotional Empathy consists of 33 items and is divided into seven intercorrelated subscales: “susceptibility to emotional contagion, appreciation of the feelings of unfamiliar and distant others, extreme emotional responsiveness, tendency to be moved by others’ positive emotional experiences, tendency to be moved by others’ negative emotional experiences, sympathetic tendency, willingness to be in contact with others who have problems” \citep{Mehrabian1972}. Similarly to the Hogan’s Empathy Scale,  QMEE is also criticised for containing items that do not necessarily test empathy, but instead emotional arousability and sympathy \citep{Eisenberg1987}. 

One of the main critiques of both these measures is that they define empathy from a single perspective, either as purely cognitive or purely emotional. An alternative, the Interpersonal Reactivity Index does not assign a single total empathy score. Instead, it has four subcategories with seven questions in each, measuring the following aspects: “perspective taking”, “empathic concern”, “personal distress”, and “fantasy”. Some theorists argue that this is the most sensible way of measuring empathy. However, \citet{BaronCohen2004} note that certain items in the IRI questionnaire, for example, “I am usually pretty effective in dealing with emergencies”, are not probing for empathy, but broader psychological characteristics. 

Finally, one of the most recent additions to the available tools that gained popularity is the Empathy Quotient \citep{BaronCohen2004}. The EQ is a self-report questionnaire for adults with average intelligence, that taps both cognitive and emotional empathy, but also covers sympathy and personal distress. It consists of 60 questions: 40 questions tapping empathy and 20 filler items to distract the participants from focusing on empathy. Each of those 40 items is scored either 0, 1, or 2 for non-empathic, mildly empathic, and strongly empathic behaviour respectively. The total maximum score is 80. Several studies have subsequently reported reliability across samples, high test-retest reliability, and concurrent validity of the EQ \citep{allison2011, lawrence2004}. 

\subsection{Factors Influencing Empathy Scores}

Previous studies have reported a variety of confounds in measuring empathy levels. For example, \textit{unemotionality} is thought to be related to low empathy \citep{frick2008}, and social desirability might influence empathy scores \citep{Duan1996}. Respondents might not indicate how they feel, but rather answer to reflect what they believe other people expect them to feel. Additionally, in the studies concerned with evaluating the effectiveness of empathy training, an individual’s level of intellect might also be a confounding factor \citep{jolliffe2004}. Another consideration is discussed by \citet{vankleef2008} in their work about the correlation between power and empathy: individuals with high social power experience less distress and compassion towards those who suffer. Finally, the literature also suggests that women tend to score significantly higher on many measures of empathic behaviour \citep{dymond1950, hogan1969, Mehrabian1972, vachon2016}. 

A large body of literature investigates the function of empathy in the context of communication between a doctor and a patient, which is a formal superior- subordinate relationship. Research suggests that empathic behaviour of a practitioner impacts significantly their professional satisfaction as well as their patient’s satisfaction. Even though these findings cannot be projected directly onto peer to peer relations, they may potentially be applied to employer-employee or manager-employee relationship in the workplace. As for social relations, evidence suggests that empathic behaviour positively impacts both interlocutors. These findings incentivise this study as they can be applied directly to the workplace relations between peers, or those at the same level in the organisational chart.

It has been demonstrated that cognitive empathy can be successfully trained by educating people, as understanding and cognition are prone to change, whereas emotion and affect are less so. However, as many researchers claim that empathy cannot be split into two separate components, this study will not delineate empathy as an exclusively cognitive characteristic but will use a questionnaire that taps cognitive aspects in order to test the effectiveness of the proposed online software on developing dispositional cognitive empathy in the individuals. Additionally, as noted by \citet{Duan1996}, the situation-specific empathic reaction is also amenable for change, hence this work will also employ a qualitative analysis of the stories, responses, and the feedback of the participants to look for evidence of a situation-specific empathic response. 

\section{Supporting Empathetic Development Through Guided Sharing of Stories}
Our aim was to ceate a supportive tool to help develop empathy amongst colleagues. The nature of the existing tools helps, but we wanted to integrate more aspects of active listening and reflection, as well as sharing, into the story experience.  We also wanted to make the interaction collaborative yet unthreatening: it needed to enage people with teach other rather than being simply another reflective tool, and yet we needed to mitigate somewhat the power imbalances within workplaces.

This led to the design ideas for ‘In Your Shoes’, a place that would embrace self-disclosure and empathic listening for exchanging personal stories. Before commencing with sketching the ideas, we established the following key principles: 

\begin{itemize}

    \item The platform is to be used by colleagues who work together daily. It will aim to assist in the creation of a more supportive and welcoming atmosphere in the workplace by establishing trust and deeper relations between the colleagues.

    \item There will be two roles in the system: a Storyteller, a person who shares their story, and a Listener, a person who reads a story. Each team member must take a turn in being a Storyteller.  Naming conventions were important --- although the Storyteller is writing things down and the Listener is, in reality, a reader of the text, by using these terms we reinforce the narrative and active listening concepts.

    \item The platform will allow participants to post stories, read, and reply to them. The intention is that reading about personal experiences, acknowledging the emotions behind the message, analysing, and reflecting should enable individuals to strengthen their listening and social skills, discover commonalities, and become more empathic towards each other.

    \item The tool will not be a live chat, but instead an asynchronous platform for posting stories and replies. Listeners will not feel compelled to reply immediately. Instead, they will have a whole day to read and thoroughly think about their response.
\end{itemize}

The asynchronous nature of the tool was a conscious decision.  There are many ways of acting synchronously: face-to-face, video-conferencing, and a lot of electronic communication tools have a semi-expectation of near synchronous communication, particularly in the workplace (Slack channels, Teams messages, WhatsApp and even email).  We wanted people to be able to reflect, digest and consider the stories that has been shared with them before replying, but equally not to have an extended period in which they may get too distracted, overtaken by other events, or simply forget. Likewise for the Storyteller, we wanted them to have space to craft their stories to reveal what they wished, and not feel pressured to produce material quickly. they also understood that the Listeners would not respond straight away and so they were slightly removed from receiving immediate responses which might have strong emotional impact.  Much longer-term communication --- letter-writing, for example --- is harder to organise in workplace settings, and so we settled on a day turnaround for the responses to stories.

\subsection{5 Days, 5 People, 5 Stories}
The design of our intervention was driven by its need to be used by working individuals.  Any use had to be short enough to achieve active daily engagement yet long enough to produce meaningful results. Based on these criteria, we designed the process to run for five days, from Monday to Friday, with five participants in each experimental group. The design of the prototype reflects this decision.

The fundamental premise of the intervention is to provide a structured environment in which people take prescribed roles of Listener or Storyteller, and through their interaction on the platform find out more about their colleagues, and develop a deeper connection and hopefully increased empathy.

\subsection{Sketching}
The first step in the design process involved creating several initial sketches to explore the fundamental principles of the system. One idea was to present different user interfaces for Storytellers and Listeners. Early sketches showed that a Storyteller could choose either a ``happy face'' or a ``sad face,'' leading to a selection of corresponding story prompts. Once logged in, a Listener would see the story along with input fields for their response, with an option for anonymity.

Feedback on these sketches raised two main points: the issue of participant anonymity and the platform for the software. This feedback led to several design decisions for the high-fidelity prototype:
\begin{itemize}
    \item The tool would be implemented as a responsive website, allowing access from multiple platforms and easier text input.
    \item Anonymity would not be allowed, as connecting stories to individuals was considered important.
    \item Story prompts would be presented immediately upon login, without categorization into positive or negative experiences.
    \item Listeners would have the option to respond, but no explicit instruction would be given to allow them to reply or not depending on how they felt.
    \item Storytellers would be able to reply to responses via the website.
\end{itemize}

\subsection{Lo-fi Prototpye}
The second step in the design process involved developing a lo-fi prototype to envision the future product in greater detail. Participants would be able to log in to the website before the activity would start. They would see their teammates and be able to familiarise themselves with the information about the activity. When the activity starts, Storytellers and Listeners would see different interfaces. Storytellers would be able to scroll through the list of emotions and choose the one they would like to talk about in their story. 

This second lo-fi prototype was shown to two people who are employed full-time in IT companies. It received chiefly positive feedback. However, it raised a discussion about the story prompts and the mechanism of replying to the Listeners. One essential aspect when designing a high- fidelity prototype was that it had to be as simple as possible. On the other hand, it also had to be robust and reflect the findings from the literature review: the solution needed to inspire participants to open up both in their stories and their replies. These two factors led to the following choices for the hi-fi prototype: 

• Storytellers and Listeners would see the same interface consisting of three pages. Firstly, the main page, Instructions, would display all the details concerning the activity. Secondly, Share a Story page would provide a free-form text input field to post a story. Finally, Read a Story page will display a story and a form for Listeners to reply to it. The decision to present the same interface would not affect the primary goal of the tool but would simplify the user experience.

• It was decided to create diversified prompts for the stories. The inspiration for the prompts came from the card game “Story Stitch” \citep{storystitch}. This game can be used for connecting people of different backgrounds through discussions. Each of the cards in this game represents a prompt to talk about, e.g., “Tell about a time when you felt proud for learning or doing something difficult”. The ideas for the story prompts were brainstormed, resulting in a range of cues.  Five random examples are: Tell about a difficult decision you had to make; Share a story when you were looked down on; Tell about your greatest passion and what it means to you; Share a story about the most important lesson learned; Share a story when you helped a stranger. 

We also removed the option to reply to the Listeners’ responses, since this was supposed to facilitate a change in understanding about the individual, and not become a conversation about the story: the person is the focus, not the narrative.  Moreover, a decision was made to not have a separate page for all the replies for a story, meaning that the only person seeing the responses was te Storyteller: this make sit more individual and provides less incentive for people to write responses for the group's consumption not the individuals. Instead, all the replies would be sent manually to Storytellers via the email.   Finally, the word ‘empathy’ was not used anywhere in the design or materials for the study, to avoid participants’ focus on empathic behaviour. 

\subsection{Final Version}
\begin{figure}
    \centering
    \includegraphics[width=1\linewidth]{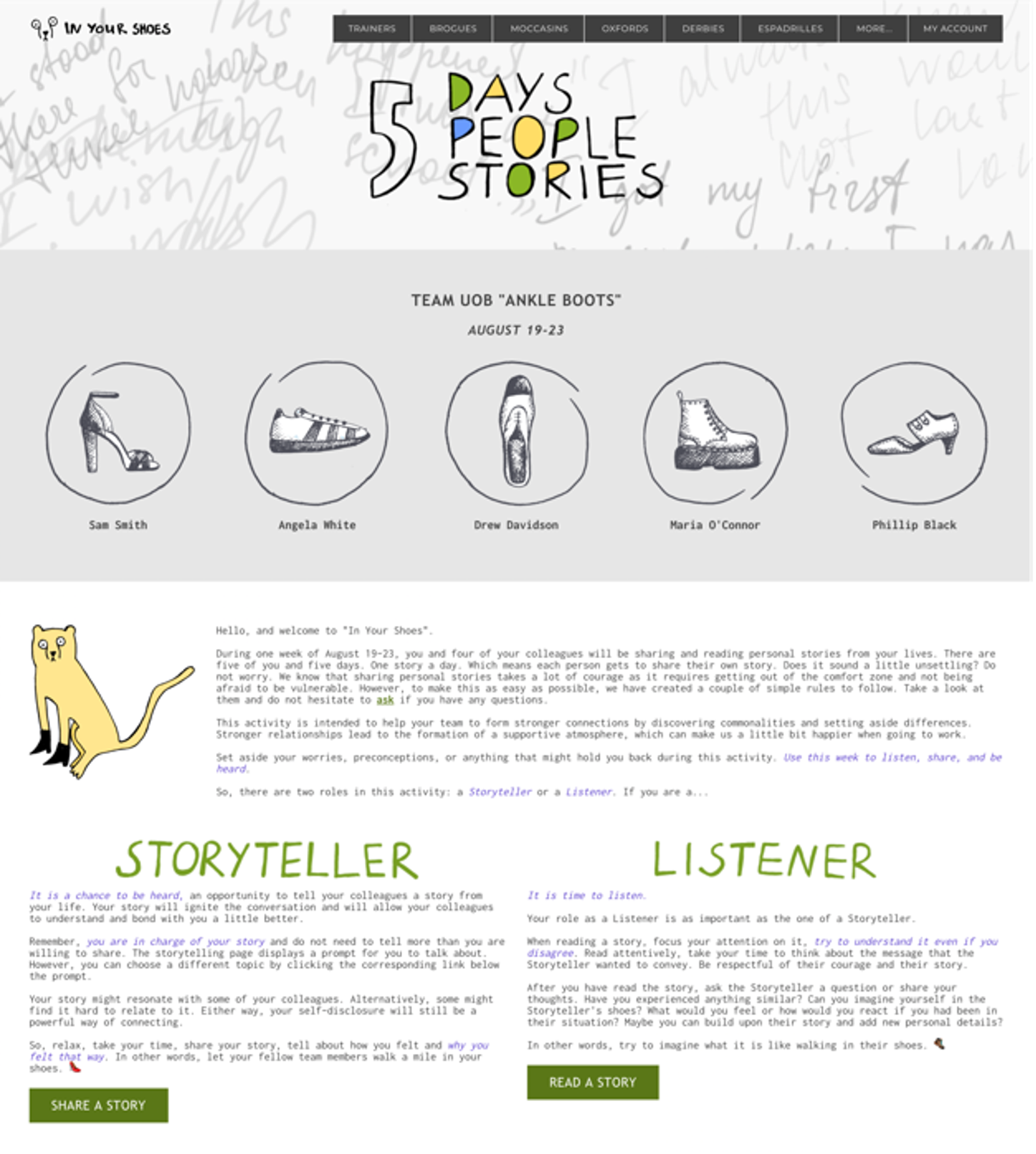}
    \caption{Design of "5 Days 5 Stories" interface}
    \label{fig:5days5storiesinterface}
    Alt text: Design of the 5 days 5 stories interface showing friendly look and feel.
\end{figure}

The final design step was to produce a working version for our exploratory study, which was then evaluated with full-time IT employees. It featured a simple, clean interface with minimal colors and hand-drawn style images to evoke a friendly, relaxed, non-corporate feel.  Pilot testing confirmed that the design met its objectives, and it was implemented on a commercial web platform to ensure data security and integrity,  site uptime and availability: it is shown in Figure \ref{fig:5days5storiesinterface}.

\section{Methodology}

The experiment was a very organic one, allowing that participants were guiding the flow of it themselves. Therefore, in order to get the most comprehensive insight into how the system affected participants’ empathy, it was decided to carry out mixed methods research. The following two research questions framed the study: 

\begin{enumerate}
    \item What effect has an online tool for exchanging personal stories on the participants’ empathy in the workplace environment?

    \item What is the participants’ experience of using the ‘In Your Shoes’ tool?

\end{enumerate}
 
\subsection{Ethical Considerations}
Care was taken during the experiment to protect participants’ personal information. Prior to commencing the study, ethical approval was sought from the University's Ethics Committee. During the recruitment process, the participants were informed about the nature of the study being about communication, what data would be collected, and how it would be manipulated and stored. They were then asked to confirm their understanding by reading and signing the consent form.  Post the study, they were informed we were looking at empathy and how to promote it, and given another chance to withdraw their data.

\subsection{Participants}
The recruitment process started four weeks before the experiment commenced. The invitation email contained a summary of the study and its potential benefits. It also had the participant information sheet attached, which explained the experiment in greater detail. Neither this email or any other ones contained word ‘empathy’ in them, instead – ‘interpersonal communication’. There was no compensation offered for participation in the study. However, the email outlined some potential personal benefits of taking part in the research. Finally, the invitation email contained a link to register for the experiment, and a form collected personal details and company name, team, and department, required for correct placement in the teams of people who work together. By the end of the recruitment phase, 33 people from six different companies, and 5 work colleagues who were also friends, had registered to participate (11 females, 26 males, one prefers not to say; 11 managerial positions, 27 non-managerial positions). Seven experimental groups were formed. Three remaining persons were not assigned to any of the groups as there was not enough interest from their companies. 

\subsection{Apparatus}
The experiment was conducted entirely remotely via email and online forms. Once a story was submitted, it was manually added to the website’s "Read a Story" page. Responses were manually collected and compiled into a single email that was sent to the Storyteller.

\subsection{Procedure}
\textbf{Before the Experiment:}  
A week prior to the experiment, participants received an email with a registration link, experiment dates, their scheduled day as Storyteller, and group names. The following day, participants were sent a link to an empathy questionnaire, which they were required to complete before the experiment commenced. A reminder email was sent on the Sunday before the experiment,  outlining the order of storytellings. 

\textbf{During the Experiment:}  
There were two roles:
\begin{itemize}
    \item \textbf{Storyteller:} The day before their designated day, participants received a reminder to prepare a story. On the day, they submitted their story via the website’s "Share a Story" page.  They were reminded that they were in charge of their narrative and were not asked to tell more than they were willing to share. The participant was encouraged to take their time and tell about what they felt and why they felt it that way. They had a selection of 16 story prompts to choose from which could either prompt their writing on the spot or be used to classify something that had pre-planned or pre-written. Later in the day or next day (if Listeners were slower to respond), the participant received the email with all the replies to their story. 

    \item \textbf{Listener:} When notified via email that a Storyteller had submitted a story, participants visited the "Read a Story" page, read the story, and were encouraged to reflect before responding, and asked to imagine their feelings and reaction had they been in the same situation. The Listener was encouraged to take their time to think about the message that the Storyteller wanted to convey, focus their attention on the story, and try to understand it even if they disagreed. They were also asked to be respectful of the Storyteller’s courage and their story. They were then invited to respond with their thoughts. 

\end{itemize}

\textbf{After the Experiment:}  
On Friday, a thank-you email was sent to all participants along with a link to complete the empathy questionnaire again. Additionally, a feedback form was provided, asking participants to detail their likes, dislikes, surprises, feelings when Storyteller and Listener, and the overall impact of the experiment on them.

\section{Results}
\subsection{Quantitative Analysis }
This study approached answering the first research question by conducting two separate analyses to assess changes in dispositional and situational empathy. Research suggests that cognition is prone to change, whereas emotions are less so. Therefore, this study chose the questionnaire that tapped the cognitive aspect of empathy to measure the participants’ empathic dispositions. Extensive research was carried out to select a suitable tool. The initial selection was made based on the number of citations of the papers introducing the tools, whether those measures were suitable for adults, were appropriate for a non-medical setting, were tapping cognitive empathy, and were available for free use. Two measures were identified: the Interpersonal Reactivity Index and the Empathy Quotient. The IRI is being criticised for including items unrelated to most conceptualisations of empathy so the Empathy Quotient was chosen. A high test-retest reliability is reported for the EQ and so it was decided to not allocate any participants to a control condition with no intervention, but instead assign all 35 individuals to the experimental condition. Two participants did not fill in the Empathy Quotient questionnaire after the experiment, resulting in 33 pairs of data points. Individual groups were too small to conduct analysis on each of them, and therefore the statistical analysis was completed on the whole group. 

\subsubsection{ Results: Dispositional Empathy}
The Shapiro–Wilk test for normality of the differences between pre- and post-intervention empathy scores yielded \(W=0.97\) with \(p=0.57\), indicating approximate normality. A paired samples t-test revealed no significant difference between pre-intervention (M = 42.4, SD = 13.6) and post-intervention (M = 41.4, SD = 14.9) scores; \(t(32) = 1.11\), \(p = 0.27\).

\subsection{Qualitative Analysis Approach} 
In order to analyse their situational empathy or their reaction to specific stories, qualitative analysis of the text was required. The experiment generated 29 stories and 109 responses. We chose one story and its four replies to conduct a qualitative analysis, which consisted of two steps. The story detaield a moment of hardship.  Firstly, sentiment analysis was used to determine emotional, language, and social tones in the corpus. Secondly, the replies were then further analysed to find evidence of situational empathy. The semantic analysis was done using the IBM Watson Tone Analyzer service \citep{ibmWatsonTone} to provide a high-level overview of the tones present in the text. This tool is widely used, and provides a clear and concise API (Application Programming Interface). This study used the general-purpose endpoint (version “2016-05- 19”) for analysing the tones of one story and four responses to it. The service returns three different categories of tones: emotion (anger, disgust, fear, joy, sadness); language (analytical, confident, tentative); and social, or Big Five personality characteristics (openness, conscientiousness, extraversion, agreeableness, emotional range). The requests were sent using Postman software.

The same four replies were then further studied to ascertain the presence of situational empathy in them. The framework that guided this analysis was adopted from the study by \citet{Coulehan2001}. In their work, the researchers discuss the stages of empathic communication between a patient and a practitioner. They suggest that these stages should be as follows: 

\begin{itemize}
    \item Active Listening, which Coulehan et al. define as non-verbal communication, including posture, eye- contact, and full attention on the speaker. 
    \item Framing or Sign Posting, or communicating the intent to listen and understand. 
    \item Reflecting the Content, or accurate identification of the content of the message and mirroring the speaker. 
    \item Identifying and Calibrating the Emotion, or identifying the emotion and adjusting the understanding about it. 
    \item Requesting and Accepting Correction, which then finalises the process once the understanding has been reached. 
\end{itemize}

\subsubsection{Results: Situational Empathy}
The Storyteller’s EQ was 64 and 65 pre and post-intervention, respectively. Their story talked about the moment of hardship and the feelings the individual had felt at that time. The sentiment analysis general-purpose endpoint returned the following interpretation of the tones in that story:
\begin{itemize}
\item Emotion: sadness (0.61), (0.63), joy (0.51)
\item Language: tentative (0.68), analytical (0.56) 
\end{itemize}

The sentiment analysis service provides scores in categories from 0-1: the higher the score the higher the likelihood that the tone is present.

This Storyteller received four replies.
\begin{itemize}
    \item Listener 1 (EQ scores: 34 and 39 pre- and post-intervention) wrote a relatively concise reply. However, they opened up in response and shared their fears had they been in the same situation. The sentiment analysis determined the following tones in their reply:
    \begin{itemize}
        \item Emotion: sadness (0.65), joy (0.70)
        \item Language: tentative (0.96), analytical (0.71)
        \item Social: agreeableness (0.91)
    \end{itemize}

    \item Listener 2 ( EQ scores: 35 and 31) imagined themselves in the Storyteller’s place and attempted to imagine the Storyteller’s feelings: “I would imagine you’d feel [...]”, “I’d be interested to know if you [...]”. The sentiment analysis produced the following result:
    \begin{itemize}
        \item Emotion: sadness (0.64), joy (0.64)
        \item Language: tentative (0.93), analytical (0.67)
        \item Social: agreeableness (0.95)
    \end{itemize}

    \item Listener 3 (EQ scores: 16 and 20) did not engage with the story but presented a similar occurrence from their life. The sentiment analysis established the following: 
    \begin{itemize}
        \item Emotion: sadness (0.61), joy (0.53)
        \item Language: tentative (0.75), analytical (0.51)
        \item Social: not scored high enough to be included in the result.
    \end{itemize}

    \item Listener 4 ( EQ scores: 63 and 60) asked several questions to clarify the Storyteller’s feelings and provided an example from their life that demonstrated their understanding. The analysis identified the following tones: 
    \begin{itemize}
        \item Emotion: sadness (0.61), joy (0.61)
        \item Language: tentative (0.83), analytical (0.75)
        \item Social: agreeableness (0.89), openness (0.86).
    \end{itemize}
\end{itemize}

All four Listeners’ emotional tones conveyed joy and sadness and were somewhat similar numerically to each other. The language tones for all Listeners were tentative and analytical. However, tone analysis of the response by Participant 3 displayed the lowest scores for the language category. Similarly, this individual was the only Listener for whom the sentiment analysis service did not identify any significant result for the social category. 

Looking at Coulehan's framework, it's recommend that the opening statements of the empathic communication should convey the desire to listen and understand. This study focuses on written texts; therefore, there was no ‘framing or sign posting’ per se. Listeners 3 and 4 thanked the Storyteller for sharing their experience. Listener 2 noted that it was a “a very interesting story”, whereas Listener 1 revealed their uncertainty regarding what their reaction should be. Two Listeners used reflection in their replies. Listener 1 agreed with the Storyteller that the circumstances described in the story were not pleasant. Listener 2 restated the idea of the story by saying: “I like how you start with [...]”, where they accurately recognised the Storyteller’s thoughts. Listeners 2 and 4 demonstrated attempt to clarify the nature of the Storytellers’ emotions. Participant 2 commented: “knowing you felt [...] allowed me to realise that it must have been [...]” or “I would imagine you’d feel [...]”. Listener 4 finished their response with a request to confirm their understanding of the idea of the story. All other Listeners’ responses did not include examples of this stage. 

What appears to be apparent is a mirroring of emotional responses overlain with an increased tentativeness --- Listeners were cautiously supportive and reflected back the issues to the Storyteller.  Broadly similar results were found across the dataset.

\subsection{Participant Feedback }

The second research question on how the people felt about participating was addressed by using grounded theory analysis on the data collected through the feedback form. Prior to analysing these data, no hypotheses were set. Grounded theory analysis let the data speak for itself to reveal significant phenomena about the participants’ experience with the software and with the activity itself. 

In total, 28 participants filled in the post-experiment feedback form. The data were exported from Google Forms and imported into the NVivo software, where the coding was carried out, which consisted of iterating through all three stages (open coding, axial coding and selective coding 
 until  saturation was reached. Analysis began as soon as the first responses started coming. As more feedback was getting submitted, the coding scheme was also adjusting to incorporate all aspects of the empirical material. 

According to the post- intervention feedback questionnaire, participants underwent a wide range of experiences during the activity well beyond empathy, including feelings of vulnerability, excitement, and safety, as well as acknowledgement of their issues. The participants’ perception of safety and personal benefit, as well as their level of enjoyment, varied depending on their role (Listener or Storyteller), the level of commitment of other participants, and the contents of other people’s stories. 

Feedback from participants revealed several themes: \textit{feeling safe; a therapeutic effect; variations in empathic behaviour; surprise and honor; feeling under pressure. }

\subsubsection{Feeling safe}
According to the participants, an important prerequisite to feeling safe was to observe other people opening up and being vulnerable. One individual stated that
\begin{quote}
“the more people reveal, the more willing and comfortable to be open it makes me”. 
\end{quote}
Another commented:
\begin{quote}
“Having read other stories before, seeing the extent to which people opened up made me feel safe to tell a story somewhat more open than I usually get”. 
\end{quote}
Two participants added that setting an example and being the first Storyteller was challenging. Perception of safety allowed participants to feel less vulnerable and exposed and encouraged them to be more open when sharing their own experiences and feelings in their stories. Additionally, safety and trust between the participants promoted opening up in the responses to the stories as well: 
\begin{quote}
“I felt that my relationship with each of the group participants allowed me to ... go more in-depth [in responses]”.
\end{quote}

\begin{figure}
    \centering
    \includegraphics[width=1\linewidth]{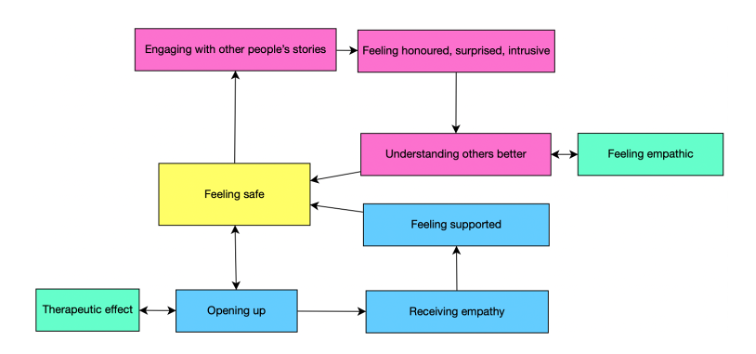}
    \caption{Participant feedback: process effect - feeling safe}
    \label{fig-safe}
\end{figure}

Figure \ref{fig-safe} displays an action sequence over time, with the perception of safety being central to both engaging with other people’s stories and sharing personal experiences. 
Alt text: Process diagram with boxes and arrows showing feeling safe as an outcome
\subsubsection{Therapeutic Effect }
One of the unexpected revelations about the experiment was its therapeutic effect on some of the participants. Eight participants mentioned they felt vulnerable and exposed when sharing: 
\begin{quote}
“it was a story from my life, and I didn’t know how others would react to it”.
\end{quote}

However, when they did talk about their experience, they noticed how rewarding it felt to submit their story, how “elevated” and “uplifting” their mood was. Additionally, nine individuals spoke about the support that they received in the replies to their stories, for example, 
\begin{quote}
“It really felt like I could feel the love”,\newline
“Everyone answered incredibly supportive and understanding of my situation and I felt much more appreciated and understood as a result of it”. 
\end{quote}

Five out of these nine participants spoke about the therapeutic effect of the activity. As one individual stated about their experience of sharing their story: 
\begin{quote}
“I knew that something had to change, but writing it out and how uncomfortable it made me feel and [...] anxious [...] made me realise that I should maybe speak to someone”. 
\end{quote}
Another person commented that
\begin{quote}
“while in a conversation I would have probably just talked about one aspect, through writing it down I was able to get a bigger picture of what I was feeling at the time of my story [...] I found it personally very therapeutic”. 
\end{quote}

\subsubsection{Empathic Behaviour }
The participants talked about empathy they felt towards others and empathy they observed from others. For example, one participant said that 
\begin{quote}
“it was nice to receive sympathy from them over my experiences”. 
\end{quote}
With regard to feeling empathic to others, experiences differed for different participants. One individual said that they
\begin{quote}
“had no empathy at all for one storyteller”. 
\end{quote}

On the other hand, one individual shared that they
\begin{quote}
“often felt quite able to put myself in the shoes of the story teller”.
\end{quote}

Strong conditions behind them supported both of these cases: existing relationship with that person
\begin{quote}
“It confirmed my thoughts about what kind of people I thought these people already were”
\end{quote}
or the degree to which one could relate to the story itself
\begin{quote}
“It is indeed easier to communicate with people who’s story resonated with me"
\end{quote}

\begin{figure}
    \centering
    \includegraphics[width=1\linewidth]{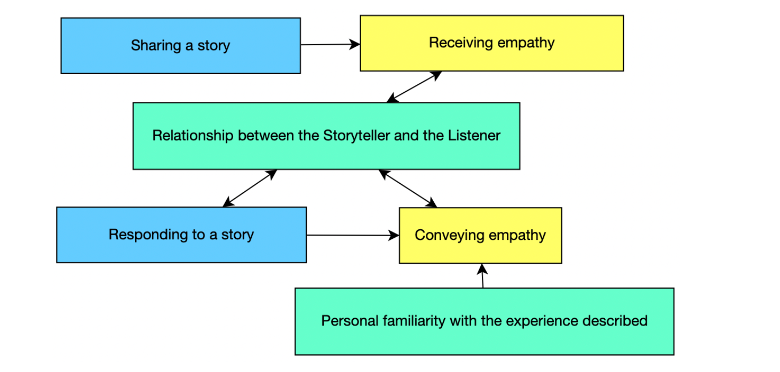}
    \caption{Participant feedback clustered under story sharing.  Process effect - empathic behavior}
    \label{fig:empathic}
\end{figure}

Figure \ref{fig:empathic} displays a chain of factors that have an impact on empathic behaviour and change throughout the activity. 
Alt text: Process diagram showing empathic behavor as an outcome
 \subsubsection{}section{Feeling Surprised or Honoured }
Thirteen participants commented on how incredibly open and personal the shared stories were. One individual said: 
\begin{quote}
“It felt people really opened up which is rare”.
\end{quote}
Two participants from one of the groups noticed that the stories were 
\begin{quote}
“getting more personal as the week goes on”.
\end{quote}
Five participants reported they felt honoured and glad to listen to such open stories. For instance, as one participant said:
\begin{quote}
“I felt happy that the people in the group felt comfortable sharing the personal stories that they did”.
\end{quote}
Five individuals were surprised by how intimate the stories were. For example, as one of the individuals said
\begin{quote}
“Some of the stories were really heart rending, I never expected my colleagues to share so much and be so personal”.
\end{quote}
Another person also mentioned how surprised they were about
\begin{quote}
“how private some people’s stories were. At least taking into account cultural norms of what people tend to talk/not talk about”. 
\end{quote}

\subsubsection{}section{Feeling under Pressure }
The final category that is worth mentioning is feeling under pressure while replying and sharing. Ten participants reported they felt anxious, vulnerable, or exposed at some point during the activity: 
\begin{quote}
“I certainly felt a little more exposed when revealing details about my life and indeed times of weakness”, \\
“I thought that nobody was going to reply because the story/I was not interesting enough or that the others had forgotten about me”.
\end{quote}
Due to the experiment’s free-flowing nature, all groups differed from each other. For example, two groups demonstrated an unexpected flow: the participants in those groups felt under pressure to impress each other both in their stories and replies. As one of the participants mentioned,
\begin{quote}
“I felt pressured to write something that others would find interesting”. 
\end{quote}
Another one said they felt
\begin{quote}
“nervous as I didn’t feel I had any interesting stories”
\end{quote}
Yet another one commented they
\begin{quote}
“felt compelled to reply given the amount of heart people have been pouring out”
\end{quote}

\subsection{Limitations}
This was an exploratory study, aimed at obtaining insights in the behavior of people when supported with this tool, and we acknowledge the relatively short duration of the investigation.  However, we also note the challenges of obtaining time from professional workers, and thank them for their deep participation and honest feedback.  We could have trialled the tool with students in a University setting, and obtained substantially more participants, but as the focus is on commercial organisations we dismissed this approach as not providing representative data, and that the six companies represented a reasonable sample of the commercial space.

The situational empathy study only focussed on one story: the personal nature of many shared meant we decided we needed to obtain further consent for analysis, out of respect for the storytellers, which was received for this story.  Anecdotally, similar results were observed across others, with the tone of responses broadly reflecting the original story. 

\section{Conclusion}
This was a pilot study investigating the impact of an online tool designed to facilitate guided interactions around narrative to facilitate both dispositional and situational empathy. Quantitative analysis of the EQ revealed no significant change in dispositional empathy over the one-week intervention. 

There are several probable explanations for this result. Firstly, one week may have been not long enough to produce any effect. Secondly, the participants were not able to have continuous conversations via the software, only one story and one reply, by design. It is unknown how many people engaged with their Listeners outside of the experiment. Perhaps, had the website functionality allowed the participants to have dialogues, it would create more opportunities for bonding, which could then lead to feeling more empathic towards each other.  We do note that participants have many other tools for ongoing conversations, however. Finally, some of the groups shared very personal and affecting stories, while a few people tried to impress each other with entertaining stories. Therefore, the contents of the stories could have been another confounding factor in this experiment. Another aspect worth considering is how the first story impacts the tone of the narratives, the direction of the experiment and consequently, the change in empathy levels in the group. 

In the case of situational empathy, one story and four replies were analysed for emotional, language, and social tones. The response of Listener 3 did not contain recognisable facets of the Big Five characteristics. This result may be correlated with the lowest EQ score in the group and the study overall, and seems to be consistent with an argument by \citet{melchers2016} which says that agreeableness is the best predictor for empathic responding. Regarding the content of the responses, all participants but Listener 3 exhibited some form of situational empathy: they reflected the message, asked for clarifications, and requested corrections. It is important to note that the results from the analysis above, undoubtedly, need to be interpreted with caution. Analysis of larger corpus is needed to affirm that semantic analysis may be useful in detecting situational empathy. 

The second question in this study sought to determine the participants’ experience of using the ‘In Your Shoes’ tool. Grounded theory analysis of the experiment feedback revealed that opening up and engaging with other people’s stories were reasonably complex processes involving multiple factors and evoking a range of various emotions. One unanticipated finding was that the ‘In Your Shoes’ appeared to have a therapeutic effect on some of the participants. The act of writing the story led to concentrating on it and possibly experiencing it all over again, which then resembled a therapy session. In their study about the benefits of reflective writing on developing empathy in clinicians, Das Gupta and Charon conclude that “the process of writing and sharing personal stories may clarify hitherto unexplored challenges or biases-vulnerabilities each physician carries with him or her throughout professional life” \citep{DasGupta2004}. They argue that practitioners’ self-awareness increases their ability to communicate empathically with their patients and thus the therapeutic effect reported by the participants suggest that empathy may indeed have been affected. 

The participants discussed a wide range of emotions and feelings they experienced during the activity. However, there was a dominant trend present: the participants spoke about the perception of safety and how it allowed them to talk about their vulnerabilities. Undoubtedly, a whole team needed to work on the creation of such an atmosphere. All participants had to be open but also had to actively listen to others and show their support. Five participants reported they were surprised by how open their colleagues were, adding that it was not in the cultural norms to open up that much. Perhaps, this mutual effort then allowed the stories to gradually become more and more personal with each day of the experiment. There exists an assumption that feeling safe is a prerequisite to empathic behaviour towards others \citep{segal2019}. Therefore, this paper supports the notion that the perception of safety reported by the participants may have possibly been a step towards more empathic behaviour among the group members. 

Whilst the limited interactions and only week-long interventions may have constrained measurable changes in dispositional empathy, the therapeutic benefits and increased interpersonal understanding suggest that the tool has potential for enhancing workplace bonding and empathy in ways not fully captured by standard metrics.

\section*{Declaration of generative AI and AI-assisted technologies in the writing process}
Generative AI has not been used in any part of the research, its design or implementation.  The paper was written by the authors. 

\section*{Declarations}

\begin{itemize}
\item Funding: none to declare
\item Conflict of interest: none
\item Ethics approval and consent to participate: all relevant ethical clearances were obtained.
\item Consent for publication: no consents required
\item Data availability: The data used in this report is highly personal and was always to be kept confidential. Therefore, regrettably the authors are unable to share the stories that generated the data.
\item Materials availability: none 
\item Code availability: none 
\item Author contribution: RB conceived the concept, advised on the intervention design, and was responsible for the final manuscript. ES designed the website for the intervention, coded it, ran the study, undertook much of the data analysis, and wrote the first draft of the manuscript.
\item Acknowledgements: We would like to thank the companies and especially the employees who gave up substantial time and, more importantly, emotional courage and commitment, in participating in this research.
\end{itemize}

\bibliography{references}
\end{document}